\documentclass[aps,prx,reprint,longbibliography]{revtex4-2}
\usepackage[T1]{fontenc}
\usepackage[utf8]{inputenc}
\usepackage{amsmath,amssymb,graphicx,bm,booktabs,microtype,xcolor}
\usepackage[colorlinks=true,allcolors=blue!55!black]{hyperref}
\hypersetup{pdftitle={Weak-measurement certification of graph-edge entanglement in PXP scar wavepackets},pdfsubject={Theoretical and numerical research article}}

\newcommand{\ket}[1]{\lvert #1\rangle}
\newcommand{\bra}[1]{\langle #1\rvert}
\newcommand{\id}{\mathbb I}
\newcommand{\Nloc}{\mathcal N^{\mathrm{loc}}}
\newcommand{\Ztwo}{\mathbb Z_2}

\begin{document}
\title{Sparse and weak-measurement certification of graph-edge entanglement in PXP scar wavepackets}
% 第一作者：王熙墨，单位 1、2
\author{Ximo Wang}
\affiliation{School of Physics and Electronic Engineering,
	Shanxi University, Taiyuan, China}
\affiliation{Collaborative Innovation Center of Extreme Optics,
	Shanxi University, Taiyuan, China}

% 第二作者：张一驰，单位 1、2
\author{Yichi Zhang}
\affiliation{School of Physics and Electronic Engineering,
	Shanxi University, Taiyuan, China}
\affiliation{Collaborative Innovation Center of Extreme Optics,
	Shanxi University, Taiyuan, China}

% 第三作者：赵茜，单位 3
\author{Xi Zhao}
\affiliation{Department of Physics,
	University of Science and Technology of China,
	Hefei, China}

% 第四作者：韩祁炜，单位 1、2
\author{Qiwei Han}
\affiliation{School of Physics and Electronic Engineering,
	Shanxi University, Taiyuan, China}
\affiliation{Collaborative Innovation Center of Extreme Optics,
	Shanxi University, Taiyuan, China}

% 第五作者：王宇航，单位 4
\author{Yuhang Wang}
\affiliation{School of Instrument Science and Opto-Electronics Engineering,
	Beijing Information Science and Technology University,
	Beijing, China}

% 第六作者：Chunxiao Du，单位 5
\author{Chunxiao Du}
\affiliation{School of Physics,
	Beihang University, Beijing, China}

% 第九作者、通讯作者：李睿，单位 8
\author{Rui Li}
\email[Contact author: ]{rli.work@buaa.edu.cn}
\affiliation{School of Applied Science,
	Beijing Information Science and Technology University,
	Beijing, China}
% 第七作者：Wenxiu Li，单位 6
\author{Wenxiu Li}
\affiliation{School of Automation (School of Artificial Intelligence),
	Beijing Information Science and Technology University,
	Beijing, China}

% 第八作者：张浩，单位 7
\author{Hao Zhang}
\affiliation{School of Space and Earth Sciences,
	Beihang University, Beijing, China}

\begin{abstract}
An imperfect many-body revival does not by itself certify the entanglement of the returning state. We give a finite-record protocol for graph-edge localizable entanglement along scar wavepackets of a graph-dressed PXP chain. The target cluster state has nonzero energy variance, so neither an exact target eigenstate nor a dark-state embedding is assumed. Fresh binary probes of the undeformed Hamiltonian's Pauli terms have a fully separable explanation, even within the dressed blockade sector. Adding local graph-stabilizer probes makes established entanglement witnesses accessible with simultaneous confidence bounds. For a specified square-root instrument, an imposed worst-case disturbance budget fixes the strength that minimizes the equal-allocation Hoeffding sufficient sampling cost. A local commutator bound accounts for finite-duration ancilla pulses while the Hamiltonian remains active. We test the protocol on chains through twenty spins, under perturbed dynamics, and with independent implementations. At the first twenty-spin return, synthetic weak records certify all nineteen graph edges. Generator, two-color, bounded-weight, and full-group benchmarks separate this localizable resource from genuine multipartite certification. The protocol measures recoverable entanglement in a known scar wavepacket, with explicit calibration and limits on its physical and statistical interpretation.
\end{abstract}
\maketitle

\section{Introduction}\label{sec:introduction}
The eigenstate thermalization hypothesis explains why local information usually becomes inaccessible during interacting quantum evolution~\cite{Deutsch1991,Srednicki1994,Rigol2008,DAlessio2016}. Quantum many-body scars provide a restricted exception: certain initial states overlap unusually strongly with atypical eigenvectors and return repeatedly inside a predominantly thermalizing spectrum. Rydberg-array experiments and the PXP model established the connection between constrained dynamics and weak ergodicity breaking~\cite{Bernien2017,Turner2018,Turner2018b}. Later theory separated approximate dynamical structures from exact eigenstate constructions, while experiments demonstrated control over scarred revivals~\cite{Serbyn2021,Moudgalya2022,Chandran2023,Ho2019,Choi2019,Khemani2019,Bluvstein2021}. This work leaves an operational question open: which entanglement properties can a sparse, weak measurement record actually certify during such evolution?

Cluster states make this question precise. Their stabilizers specify a graph of entangling correlations, and their local measurements underpin measurement-based quantum computation~\cite{Briegel2001,Raussendorf2001,Raussendorf2003,Hein2004}. Entanglement localized onto two selected vertices is an operational resource distinct from the entropy of an unmeasured subsystem~\cite{Verstraete2004,Popp2005}. Local stabilizer witnesses connect this resource to measurements on the original many-body state, without full tomography or physically localizing entanglement on every diagnostic copy~\cite{Amaro2018Witnesses,Amaro2020Localizable}. Economical cluster-state witnesses, multipartite structure tests and restricted-measurement criteria are also well developed~\cite{Toth2005,Toth2005b,Guhne2009,Zhou2019,Li2026}. The question here is how to combine these established witnesses with a specified weak instrument, a limited per-copy disturbance, and simultaneous finite-record guarantees along an approximately recurrent many-body trajectory.

Exact scar embeddings address a different question: how can an exceptional state be made an exact eigenvector of a nonintegrable Hamiltonian? General embedding constructions, exact PXP states and exact towers answer this in several ways~\cite{Shiraishi2017,Lin2019,Moudgalya2018,Schecter2019,Mark2020}. Stabilizer-scar Hamiltonians, single-scar observation protocols and decoherence-free scar sectors connect these ideas to structured entanglement and protection~\cite{Hartse2025,Dooley2026,Larsen2026,Wang2024}. A filtering argument that conserves an exact dark-state population does not automatically extend to an approximately recurrent state with a nonzero energy variance. A no-click history can itself reflect measurement backaction rather than the entanglement of the unmonitored state. A recent preprint uses exact stabilizer-scar subspaces and direct fidelity estimation to benchmark nonequilibrium quantum simulations under specified noise assumptions~\cite{Hartse2026Benchmark}. Here we consider imperfect returns of a graph-dressed PXP wavepacket, whose target graph state is explicitly not an energy eigenstate.

Other characterization methods address complementary resource regimes. Direct fidelity estimation evaluates selected expectation values; compressed and matrix-product-state tomography use promises about the input family~\cite{Flammia2011,DaSilva2011,Gross2010,Cramer2010}. Classical shadows and derandomized Pauli measurements predict many observables from randomized records~\cite{Huang2020,Huang2021}. Randomized measurement protocols also give access to entropic moments and entanglement signatures~\cite{Elben2018,Brydges2019,Elben2019,Vermersch2018,Ketterer2019}. Verification theory provides rigorous soundness guarantees in stabilizer, nondemolition and adversarial settings, among others~\cite{Pallister2018,Zhu2019,Dangniam2020,Liu2021,Riera2023,Yu2019,Zhang2020,Bennink2021}. Counting distinct settings alone does not suffice to compare these methods: a global product-basis shot, a local ancilla parity check and a continuous measurement trajectory require different operations and disturb different degrees of freedom.

We study this question in a physically graph-dressed PXP chain. The dressing carries a known approximate scar wavepacket into a frame with a prescribed cluster target. Fresh binary readouts of every Pauli term in the undeformed Hamiltonian are blind to the relevant entanglement: we show that the entire menu admits a fully separable explanation, including within the dressed blockade sector. We then add randomly sampled graph stabilizers to the accessible menu. Established localizable-entanglement witnesses applied to this record give graph-edge negativity certificates with explicit sampling and calibration errors. For the chosen square-root instrument, an imposed worst-case disturbance budget fixes the largest admissible strength and hence the smallest Hoeffding sufficient preparation budget within an equal-allocation protocol. Finite-duration ancilla pulses can be included while the Hamiltonian remains active. Simulations through $N=20$ test the certificates, and separate global-entanglement benchmarks identify which stronger conclusions require additional observables. The Supplemental Material contains detailed derivations and reproducibility records~\cite{SupplementalMaterial}.

\section{A PXP wavepacket and an uninformative measurement menu}\label{sec:model}
\subsection{A graph-dressed PXP wavepacket}
Consider an open chain of $N$ qubits. Sites are indexed from $0$ to $N-1$, $n_i=(\id-Z_i)/2$, $P_i=\id-n_i$, and missing boundary projectors are identities. The reference Hamiltonian is
\begin{equation}
 H_0=J\sum_{i=0}^{N-1}P_{i-1}X_iP_{i+1},\qquad J>0.
 \label{eq:pxp}
\end{equation}
The dynamics remain in the invariant blockade sector with $n_in_{i+1}=0$, of dimension $F_{N+2}$, where $F_k$ is a Fibonacci number. Statements about the nonintegrable spectral background refer to this sector. We do not pool it with the other fragmented sectors of the unconstrained operator.

Let $h=(X+Z)/\sqrt2$ and define
\begin{equation}
 W=\Big(\prod_{i=0}^{N-2}\mathrm{CZ}_{i,i+1}\Big)h^{\otimes N},\quad
 K_i=X_i\prod_{j:\,|j-i|=1}Z_j .
 \label{eq:clifford}
\end{equation}
The product includes only existing neighbors. Since $WZ_iW^\dagger=K_i$ and $WX_iW^\dagger=Z_i$, the physical Hamiltonian $H_G=WH_0W^\dagger$ is
\begin{equation}
 H_G=J\sum_i\frac{\id+K_{i-1}}2\, Z_i\,
                      \frac{\id+K_{i+1}}2 .
 \label{eq:dressed}
\end{equation}
Absent boundary factors in this equation are identities. Each summand has support within five consecutive sites, and its factors commute with each other. This gives an exact local change of representation of a known constrained model; it provides no evidence for a new universality class.

For even $N$, let $\ket{\Ztwo}$ have occupied even sites and empty odd sites, set $s_i=(-1)^{(\Ztwo)_i}$, and write
\begin{equation}
 \begin{aligned}
 \ket{G_s}&=W\ket{\Ztwo},\qquad A_i=s_iK_i,\\
 \ket{\psi_G(t)}&=We^{-iH_0t}\ket{\Ztwo}.
 \end{aligned}
 \label{eq:trajectory}
\end{equation}
The $A_i$ stabilize the initial signed cluster state. The experimentally accessible means $a_i(t)=\langle A_i\rangle_t$ equal $s_i\langle Z_i\rangle_t$ in the reference frame. We use this equality to accelerate simulation; the certification itself concerns the physically dressed state. Bare $Z$ measurements alone do not certify entanglement in the undressed system.

The graph state is explicitly not an eigenstate:
\begin{equation}
 \langle H_G\rangle_{G_s}=0,\qquad
 \operatorname{Var}_{G_s}(H_G)=\frac{N J^2}{2}>0.
 \label{eq:variance}
\end{equation}
Each occupied N\'eel site produces a distinct orthogonal flipped configuration, which proves Eq.~\eqref{eq:variance}. No annihilator is added to make this vector dark. The finite-depth dressing preserves the spectrum and changes the pure-state entanglement entropy across a contiguous spatial cut by at most a size-independent constant. It therefore carries approximate scarred dynamics into a graph-entangled physical frame without turning the wavepacket into an embedded eigenstate.

\begin{figure*}[t]
\centering\includegraphics[width=\textwidth]{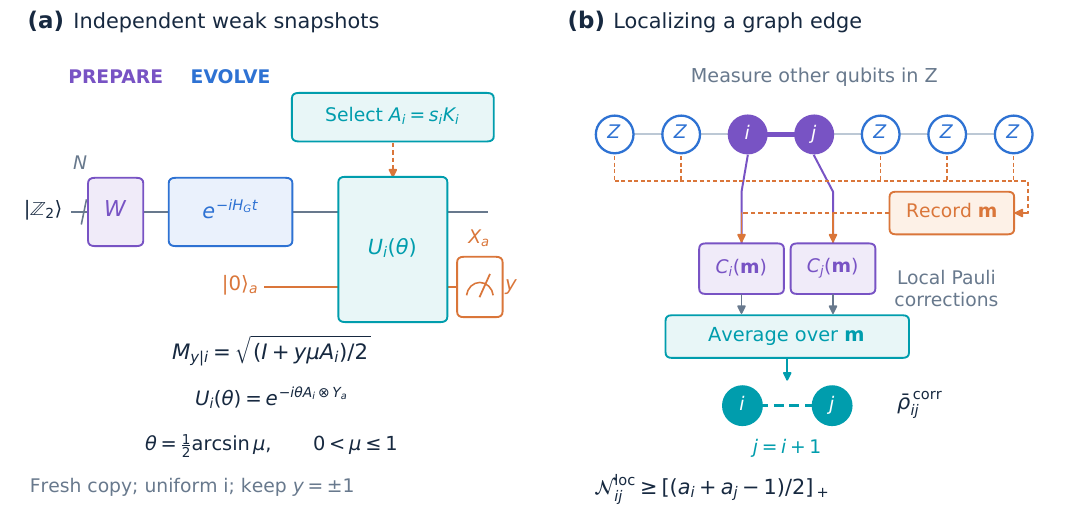}
\caption{Graph-edge certification from independent weak probes. (a) Each shot prepares $W\ket{\Ztwo}$, evolves to a nominated time, selects one local signed stabilizer $A_i$, and retains either calibrated binary outcome. Only one parity is probed per copy. (b) Measuring the other vertices in $Z$ and applying known endpoint corrections gives a constructive localizable-negativity bound; the diagnostic copies need not undergo this localization procedure.}
\label{fig:protocol}
\end{figure*}

\subsection{Why Hamiltonian-term sampling alone is insufficient}
Expand $H_0$ into Pauli summands, each containing one $X_i$ and zero, one or two neighboring $Z$ operators. Let $C_0=\prod_iZ_i$. The Hamiltonian and each summand are real and change $C_0$ parity. Evolution from a real computational-basis vector gives real amplitudes in its initial parity sector and purely imaginary amplitudes in the other, up to a common phase. Every real parity-changing Hermitian summand thus has exactly zero expectation. Clifford conjugation gives the same result for every Pauli summand $B_\ell$ of $H_G$:
\begin{equation}
 \langle B_\ell\rangle_t=0 \quad\text{for all }\ell,t.
 \label{eq:blindness}
\end{equation}
For binary fresh-copy readouts with effects $(\id+y\mu B_\ell)/2$, all outcomes are therefore fair coins. The fully mixed state $\id/2^N$ is fully separable and produces the same distributions. This obstruction persists under a snapshot promise of dressed-blockade support. To see this, fix the even reference sites in $\ket0$ and take the odd sites maximally mixed. The dressed image of this mixture lies in $W\mathcal H_{\mathrm{bl}}$ and is fully separable: choose independent uniform $Z$ eigenvalues on the odd physical sites and prepare each even site in the $X$ eigenstate whose sign is the product of its neighbors' $Z$ values. Averaging these product states gives exactly the dressed reference mixture. The reference mixture is diagonal, so its Pauli-term means are likewise zero. This menu cannot certify snapshot entanglement, regardless of repetitions (Fig.~S1(a) of the Supplemental Material~\cite{SupplementalMaterial}). The comparison does not assume a separable trajectory generated by the same trusted Hamiltonian from the prescribed pure initial state; that stronger dynamical promise poses a different inference problem. The result concerns the specified menu and instrument, not all measurements derivable from $H_G$. In particular, products of terms used to estimate $H_G^2$ carry additional information. A deformation with diagonal terms also need not obey Eq.~\eqref{eq:blindness}.

\section{Local certification under a disturbance budget}\label{sec:certificate}
\subsection{Local weak readout with finite-record guarantees}
Choose $i$ uniformly at random on each fresh snapshot and perform
\begin{equation}
 M_{y|i}=\sqrt{\frac{\id+y\mu A_i}{2}},\quad y=\pm1,
 \quad 0<\mu\leq1 .
 \label{eq:instrument}
\end{equation}
Writing $\theta=\tfrac12\arcsin\mu$ makes both the Kraus operator and its Born probabilities explicit:
\begin{equation}
 \begin{aligned}
 M_{y|i}&=\frac{\cos\theta\,\id+y\sin\theta\,A_i}{\sqrt2},\\
 p(y|i)&=\tfrac12(1+y\mu a_i),\qquad \mathbb E[y|i]=\mu a_i.
 \end{aligned}
 \label{eq:weakBorn}
\end{equation}
Because $A_i=A_i^\dagger$, $A_i^2=\id$ and $0<\theta\leq\pi/4$, these are the positive square roots of the stated effects, and the effects sum to the identity. Here ``weak'' means partial information at finite strength, with no anomalous weak value or postselection~\cite{Aharonov1988,Dressel2014}. The square-root instrument and its implementation both matter: measuring and resolving the constituent one-qubit Paulis separately produces a different disturbance~\cite{Liu2021,Brun2002,Jacobs2006}.

After $M$ shots at one time, let $m_i$ denote the actual number assigned to site $i$ and, for $m_i>0$, define $\widehat a_i=(m_i\mu)^{-1}\sum_{r:i_r=i}y_r$. For independent ideal shots on identical inputs,
\begin{equation}
 \begin{aligned}
 \mathbb E[\widehat a_i\mid m_i]&=a_i,\\
 \operatorname{Var}(\widehat a_i\mid m_i)&=
          \frac{1-\mu^2a_i^2}{m_i\mu^2}.
 \end{aligned}
 \label{eq:weakEstimatorVariance}
\end{equation}
An unbiased estimate from a finite record can lie outside $[-1,1]$; the physical mean still lies within the observable's spectrum. At regular parameter points $\mu|a_i|<1$, the classical Fisher information about $a_i$ in one binary shot is
\begin{equation}
 \mathcal I_i(a_i)=\frac{\mu^2}{1-\mu^2a_i^2}.
 \label{eq:weakFisher}
\end{equation}
In the weak limit, information per copy therefore falls as $\mu^2$; fixed accuracy cannot be obtained at vanishing preparation cost.

For a fixed family of $T$ times and $0<\alpha<1$, define
\begin{equation}
 r_i=\sqrt{\frac{2\log(2NT/\alpha)}{m_i\mu^2}},\qquad
 a_i^- =\max\{-1,\widehat a_i-r_i-b_i\}.
 \label{eq:interval}
\end{equation}
For an unobserved site, set $r_i=\infty$ and $a_i^-=-1$. The term $b_i$ bounds calibrated systematic readout bias. With ideal effects, conditioning on the full independently chosen setting schedule $\mathcal A$ gives Hoeffding's inequality, for $u>0$ and $m_i>0$,
\begin{equation}
 \Pr\!\left(|\widehat a_i-a_i|\geq u\mid\mathcal A\right)
 \leq2\exp\!\left(-\frac{m_i\mu^2u^2}{2}\right).
 \label{eq:conditionalTail}
\end{equation}
At $u=r_i$, each failure probability is at most $\alpha/(NT)$. A union bound over sites and times, followed by averaging over schedules, gives simultaneous coverage of at least $1-\alpha$~\cite{Hoeffding1963}. If the calibrated mean differs from $a_i$ by at most $b_i$, the same inequality first bounds fluctuations about that actual mean. Adding $b_i$ then gives $a_i\geq a_i^-$ on the same confidence event. This does not require all $a_i$ to be equal. Under uniform site sampling, the mean of $y/\mu$ separately estimates the spatial mean. Averaging over the setting distribution proves this directly; it is an inverse-probability weighting identity, distinct from the without-replacement setting of Ref.~\cite{Horvitz1952}. The spatial mean alone cannot certify an unmeasured worst-case edge.

The confidence statement assumes independent preparations, calibrated effects, and times fixed independently of the outcomes. A deterministic bias bound $b_i$ enters through the triangle inequality; a statistically estimated calibration bound needs its own failure budget. Conditioning on all settings preserves the independent outcome law, so the random counts in Eq.~\eqref{eq:interval} need not be replaced by their mean. For independent but nonidentically prepared inputs, the concentration argument bounds the average mean of those inputs; any deviation of that average from the nominated snapshot must also be bounded. The argument does not cover unknown preparation correlations, adaptive stopping or successive backaction-dependent readouts of one system. Each copy supplies one local parity, and coverage of all sites accumulates across copies; unseen graph structure is not inferred from a fixed small subset.

\subsection{Established witnesses and the local certificate}
We use the witness-based localization framework of Amaro, M\"uller, and Pal~\cite{Amaro2018Witnesses,Amaro2020Localizable}, together with the neighboring-stabilizer threshold of T\'oth and G\"uhne~\cite{Toth2005b}. For any input density matrix, measure all vertices except neighboring $i,j=i+1$ in $Z$, then correct the endpoint Pauli byproducts and the known target signs. The averaged corrected two-qubit state has graph-Bell stabilizer expectations $a_i$ and $a_j$. Its fidelity with the two-qubit graph state satisfies $F_{ij}\geq(a_i+a_j)/2$. We use the partial-transpose negativity $\mathcal N(\sigma)=(\|\sigma^{T_j}\|_1-1)/2$~\cite{Vidal2002,Peres1996}. The Bell projector has partial-transpose operator norm $1/2$, so the trace inequality gives $\mathcal N(\sigma)\geq F_{ij}-1/2$. On the simultaneous confidence event, this gives the following certificate, valid also for mixed states:
\begin{equation}
 \Nloc_{ij}(\rho)\ \geq\
 \left[\frac{a_i+a_j-1}{2}\right]_+
 \ \geq\ \left[\frac{a_i^-+a_j^--1}{2}\right]_+ .
 \label{eq:edge}
\end{equation}
Here $[x]_+=\max(0,x)$ and $\Nloc$ is the maximum average two-qubit negativity achievable by allowed local measurements on the other vertices. The specified $Z$-measurement procedure gives a lower bound; we do not solve the full optimization. This quantity differs from the negativity of the unconditioned two-site reduced state and from concurrence~\cite{Wootters1998}. Our normalization is one half of the trace-norm excess used in parts of the earlier localization literature~\cite{Amaro2018Witnesses}.

Using only two single-generator means gives a conservative specialization of an existing projector witness. If the additional product mean $c_{ij}=\langle A_iA_j\rangle$ is accessible, the corrected graph-Bell projector has fidelity $(1+a_i+a_j+c_{ij})/4$, giving the stronger localizable-negativity lower bound $[(a_i+a_j+c_{ij}-1)/4]_+$. The commuting-generator inequality $c_{ij}\geq a_i+a_j-1$ reduces this bound to Eq.~\eqref{eq:edge}. Equation~\eqref{eq:edge} thus uses less information in exchange for a smaller menu; it introduces no new localization principle. For general-state certification, separate single-generator shots do not determine the product means. A trusted exact dressed-blockade support promise gives additional identities, discussed in Sec.~S14 of the Supplemental Material; the local certificate itself needs no such promise.

Each positive edge certifies entanglement across any bipartition separating its endpoints. Certifying every edge of a connected graph therefore rules out separability across every fixed cut. For mixed states, this is full inseparability; it does not exclude all mixtures of states separable across different cuts. Genuine multipartite entanglement (GME) excludes that convex mixture as well. The distinction matters operationally: mixing, over all sites, the target state completely dephased in $Z$ on one chosen site gives a biseparable state with $a_i=1-1/N$, so every edge in Eq.~\eqref{eq:edge} is positive for $N>2$. All-edge positivity therefore does not establish mixed-state GME.

A separate generator inequality bounds the fidelity with the entire target:
\begin{equation}
 F_G=\bra{G_s}\rho\ket{G_s}\geq
 1-\sum_i\frac{1-a_i}{2}\equiv L_G .
 \label{eq:global}
\end{equation}
This is the standard generator fidelity bound~\cite{Toth2005b}. In the common eigenbasis, $\sum_i(1-A_i)/2$ counts defects and is at least one outside the target subspace; its expectation proves Eq.~\eqref{eq:global} without a purity or blockade promise. For a connected graph, each nontrivial cut has a nonzero binary cross-adjacency matrix. The graph-state Schmidt weights are all $2^{-r}$ for a binary rank $r\geq1$. The maximal squared overlap with a product state across any cut is therefore at most $1/2$, and the same bound holds for mixtures over cuts. Thus $F_G>1/2$ certifies GME; using the lower endpoints in Eq.~\eqref{eq:global} gives a valid finite-record sufficient test~\cite{Hein2004,Toth2005b}. Failure of this test does not show that GME is absent. As size increases, summing local deficits makes this particular global bound more demanding. More advanced restricted-measurement criteria can improve certification and complement the weak-instrument analysis~\cite{Li2026}.

When only this generator fidelity bound is needed, uniform site sampling can estimate the spatial mean $\bar a=N^{-1}\sum_i a_i$ directly at one nominated time. With $0<\alpha_{\rm av}<1$,
\begin{equation}
 \widehat{\bar a}=\frac{1}{M\mu}\sum_{r=1}^{M}y_r,\qquad
 r_{\rm av}=\sqrt{\frac{2\log(2/\alpha_{\rm av})}{M\mu^2}}
 \label{eq:spatialMean}
\end{equation}
gives an ideal-instrument interval for the mean with confidence $1-\alpha_{\rm av}$. Since $L_G=1-N(1-\bar a)/2$, its statistical half-width is $Nr_{\rm av}/2$. Requiring this half-width to be at most $\epsilon_F>0$ gives the sufficient count
\begin{equation}
 M\geq\frac{N^2}{2\mu^2\epsilon_F^2}\log\frac{2}{\alpha_{\rm av}}.
 \label{eq:meanGlobalBudget}
\end{equation}
This avoids adding separate sitewise errors when only $L_G$ is needed, but cannot certify an unmeasured worst-case edge. If both statements are requested, the failure budget $\alpha_{\rm av}$ is separate from the site--time family; this scaling is not a lower bound for all graph-state verification protocols.

\subsection{Why use a weak probe: an explicit disturbance constraint}
For the specified square-root instrument, adding the two unnormalized conditional outputs gives
\begin{equation}
 \mathcal E_i(\rho)=(1-q)\rho+q A_i\rho A_i,\qquad
 q=\frac{1-\sqrt{1-\mu^2}}2 .
 \label{eq:channel}
\end{equation}
With trace distance $D(\rho,\sigma)=\|\rho-\sigma\|_1/2$,
$D(\mathcal E_i(\rho),\rho)=qD(A_i\rho A_i,\rho)\leq q$.
An equal superposition of $+1$ and $-1$ eigenvectors of $A_i$ saturates the bound. This bound applies to the outcome-averaged state, not to every conditioned branch. The exact relation $\mu^2=4q(1-q)$ and the small-strength expansion $q=\mu^2/4+O(\mu^4)$ connect disturbance with estimator variance and Fisher information. Figure~S1(b) of the Supplemental Material shows the ideal disturbance curve~\cite{SupplementalMaterial}.

Impose a worst-case disturbance budget $0<\delta<1/2$ on each ideal probe. For this instrument family the budget is exactly $q\leq\delta$, giving
\begin{equation}
 \mu^2\leq4\delta(1-\delta).
 \label{eq:disturbancebudget}
\end{equation}
With equal fixed counts per site at one time, a total preparation budget sufficient for simultaneous statistical accuracy $\epsilon_{\mathrm{stat}}>0$ with probability at least $1-\alpha$ is
\begin{equation}
 M\geq\frac{2N}{\mu^2\epsilon_{\mathrm{stat}}^2}
          \log\frac{2N}{\alpha}.
 \label{eq:resources}
\end{equation}
The bound decreases monotonically with $\mu$. The largest allowed strength, $\mu_\delta=2\sqrt{\delta(1-\delta)}$, therefore minimizes the sufficient preparation budget for the stated fixed-allocation protocol:
\begin{equation}
 M_{\mathrm{suff}}(\delta)=
 \frac{N}{2\delta(1-\delta)\epsilon_{\mathrm{stat}}^2}
       \log\frac{2N}{\alpha}.
 \label{eq:budgetcost}
\end{equation}
The right side gives a continuous budget; an executable allocation uses $N$ times the ceiling of the corresponding per-site count. Uniform random allocation requires either checking the realized $m_i$ or allowing for its allocation tail. For $0<\eta<1$ at one nominated time, the tail obeys
\begin{equation}
 \Pr\!\left\{\min_i m_i<(1-\eta)M/N\right\}
 \leq N e^{-\eta^2M/(2N)}.
 \label{eq:allocationTail}
\end{equation}
To obtain a statistical radius at most $\epsilon_{\rm stat}$ with an additional allocation-failure allowance $0<\beta<1$, choose an integer $M$ satisfying both
\begin{equation}
 \begin{aligned}
 M&\geq\frac{2N}{(1-\eta)\mu^2\epsilon_{\rm stat}^2}
           \log\frac{2N}{\alpha},\\
 M&\geq\frac{2N}{\eta^2}\log\frac{N}{\beta}.
 \end{aligned}
 \label{eq:randomAllocationBudget}
\end{equation}
The planned precision and simultaneous coverage then hold with joint failure probability at most $\alpha+\beta$. Intervals computed from the actual counts retain their $1-\alpha$ coverage even if the planned precision is not reached. Coverage and precision are separate requirements: a realized interval can have the stated coverage before every site has received enough samples.

Equation~\eqref{eq:budgetcost} sets no lower bound over all estimators, instruments, collective measurements, or state-dependent disturbance criteria. It fixes a statistical error target for the ideal instrument. Finite-pulse bias and total precision need further constraints, given in Sec.~S13 of the Supplemental Material~\cite{SupplementalMaterial}. Sections S3 and S4 give the full instrument and conditional-probability derivations.

A disturbance constraint is useful when hardware limits backaction or the output is reserved for a later operation. We do not simulate that later use. If each measured copy can be discarded and no disturbance budget applies, $\mu=1$ gives the smaller sufficient preparation cost. Weak probes are therefore a controlled resource choice, with no intrinsic sample advantage. Full-register product-basis and shadow protocols use different per-copy resources and are not ruled out by this comparison.

\section{Finite-size dynamics and entanglement benchmarks}\label{sec:benchmarks}
\begin{figure*}[t]
\centering\includegraphics[width=\textwidth]{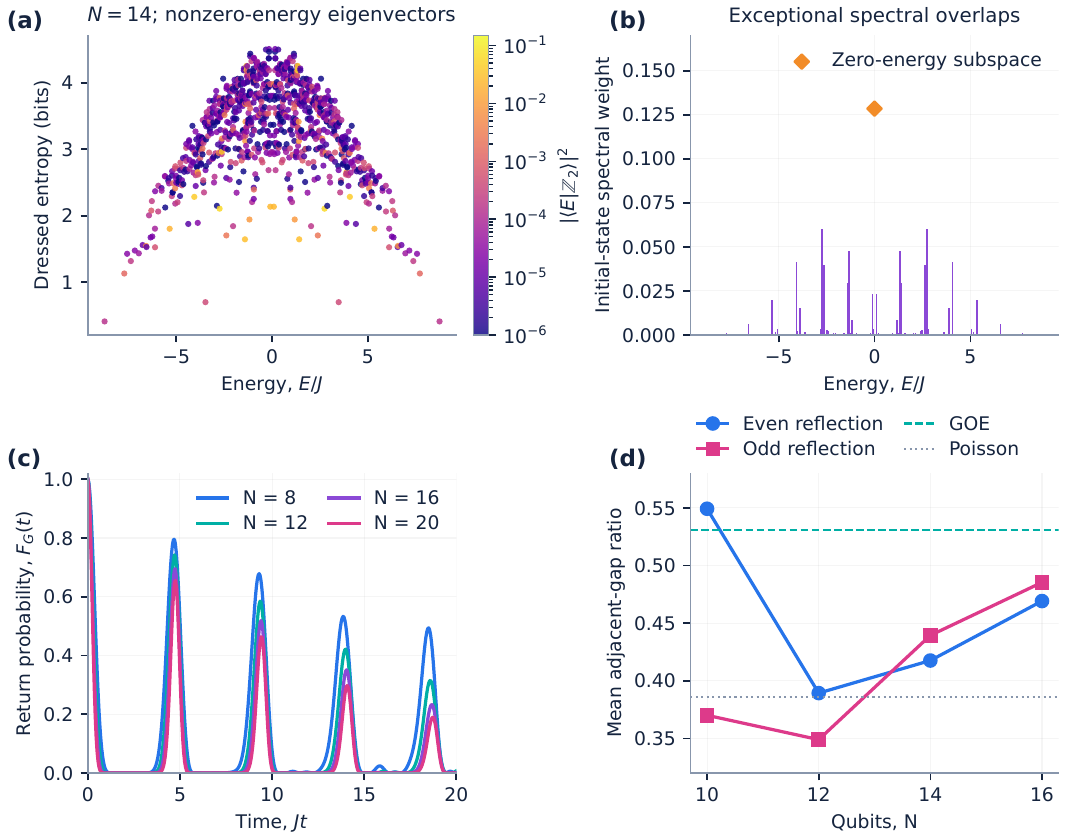}
\caption{Finite-size evidence connecting the certified trajectory to PXP scarring. (a) Dressed half-chain entropies of all nonzero-energy eigenvectors for $N=14$, colored by their N\'eel overlaps. Both reflection sectors are shown for visualization; they are resolved separately for statistics. (b) Spectral overlap distribution, with the entire degenerate zero-energy contribution shown as one diamond to avoid a basis-dependent zero-mode label. (c) Exact sparse propagation for $N=8,12,16,20$ shows imperfect recurrent returns. (d) Central-spectrum gap ratios within each reflection sector, with zero modes and adjacent triplets crossing their removed interval excluded. Finite-size values fluctuate substantially; the GOE and Poisson lines are reference values, not fitted asymptotes. This deterministic size series has no disorder confidence intervals. Spectral outliers and known PXP dynamics support the scar interpretation; the data do not prove thermodynamic ETH.}
\label{fig:scars}
\end{figure*}

\subsection{Spectral structure and recurrent graph-edge entanglement}
We propagate the constrained state exactly using sparse matrix-exponential action for $N=8,10,\ldots,20$, and resolve both reflection sectors for spectra up to $N=16$ (Sec.~\ref{sec:numerics}). The dressed spectral entropy and N\'eel overlap reveal exceptional eigenvectors across the spectrum, alongside imperfect recurrent quench returns (Fig.~\ref{fig:scars}). We treat exact zero modes as degenerate subspaces; their basis-dependent overlaps do not identify individual scar vectors. Adjacent-gap ratios give supporting finite-size evidence: at $N=16$ the even and odd sectors yield $0.4692$ and $0.4854$, below the GOE reference $0.5307$~\cite{Oganesyan2007,Atas2013}. Large fluctuations at smaller sizes prevent a claim of convergence to a random-matrix limit.

At the first $N=20$ return maximum on the $0.05/J$ grid, $Jt=4.75$, the exact return fidelity is $0.655679$ and the weakest-edge negativity bound is $0.271313$. By comparison, $L_G=-0.080433$ in Eq.~\eqref{eq:global}, so its physical lower bound is only zero. Exact fidelity above $1/2$ would establish GME if independently certified. Here it serves as a simulation reference, not a result inferred from the local records. We show both quantities to keep this distinction explicit.

\begin{figure*}[t]
\centering\includegraphics[width=\textwidth]{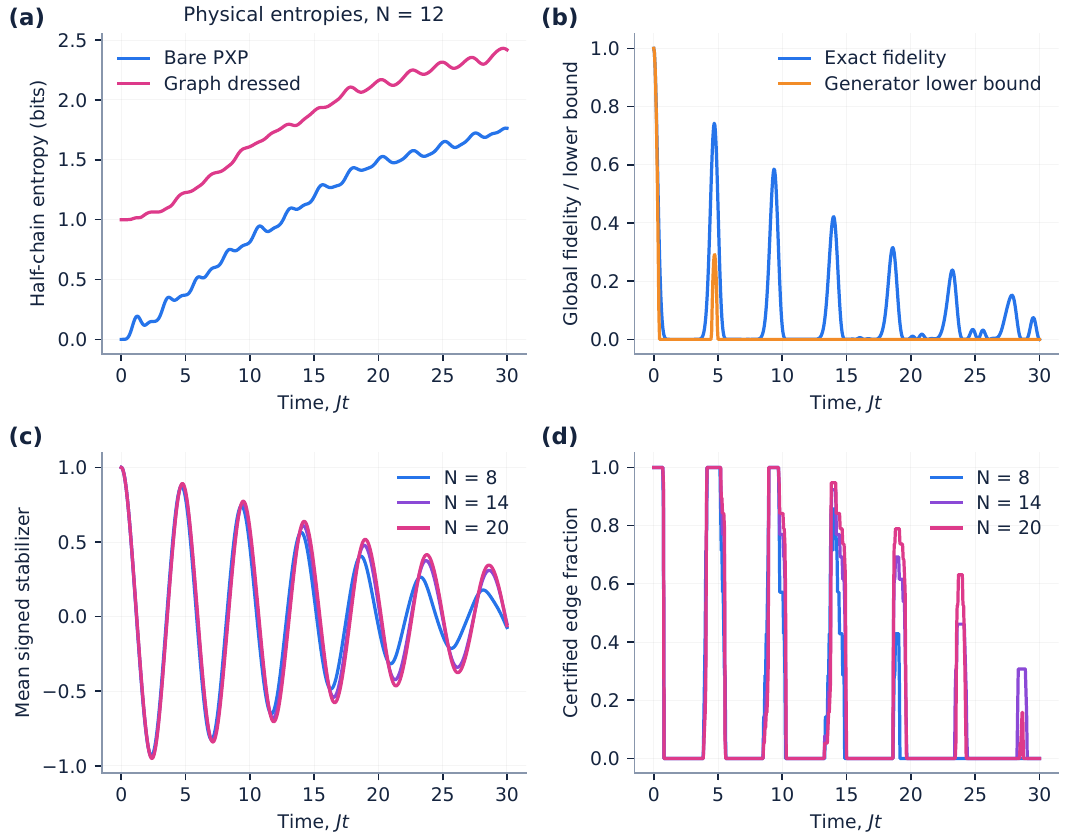}
\caption{Additional dynamical diagnostics across the full simulated window. (a) Bare and physically graph-dressed half-chain entropy for $N=12$, both computed from state singular values in bits. The finite-depth circuit changes the entropy but preserves return probabilities. (b) Exact target fidelity and the generator-based global lower bound for the same chain, with negative values of the latter clipped at zero. (c) Mean signed stabilizer for three sizes. The negative part of an oscillation corresponds to a different graph-syndrome pattern and is not interpreted as absence of all entanglement. (d) Fraction of graph edges whose exact localizable-negativity sufficient bound is positive. These are noiseless computed witness values; Fig.~\ref{fig:certification} additionally includes finite-record confidence.}
\label{fig:sextended}
\end{figure*}

Figure~\ref{fig:sextended} shows entropy, fidelity, signed-stabilizer and exact edge-bound diagnostics for the full window. The noiseless dynamical references are separate from the finite-record certificates below.

\subsection{Simultaneous certificates from synthetic weak records}
\begin{figure*}[t]
\centering\includegraphics[width=\textwidth]{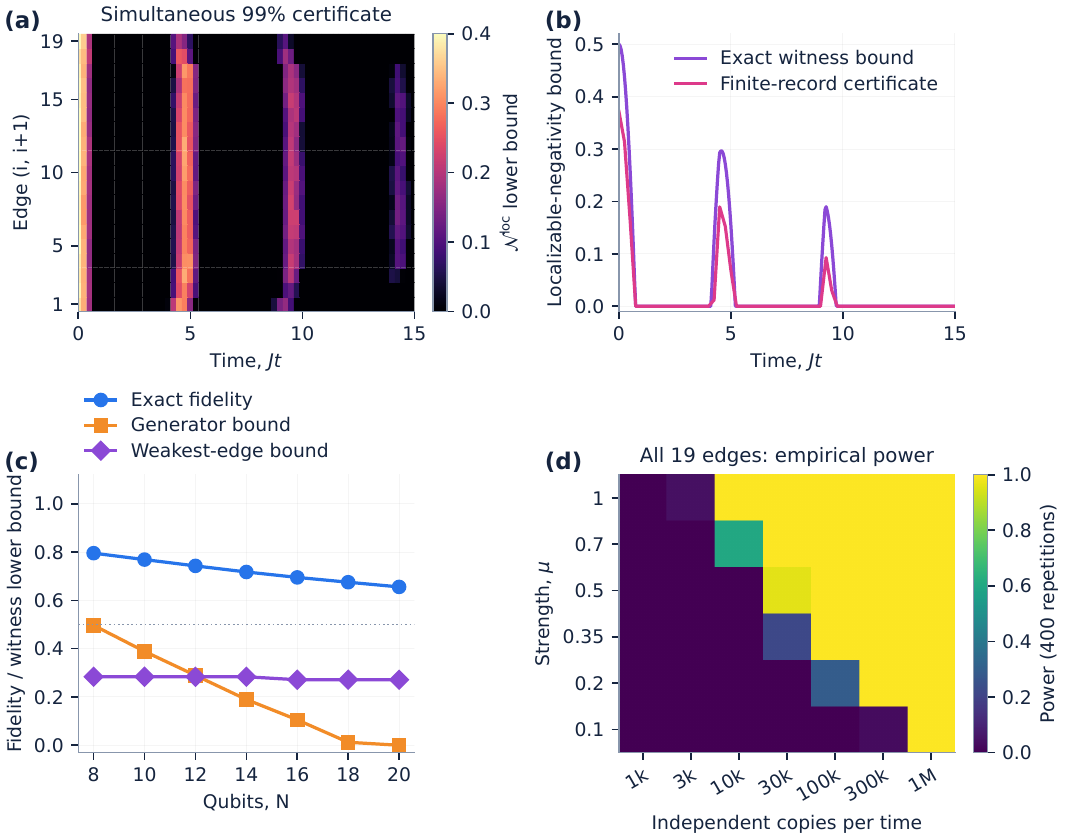}
\caption{Cluster-edge certification from spatially sparse weak records. (a) Lower bounds on localizable negativity on all 19 edges of the $N=20$ chain, using one check per copy, $\mu=0.5$ and $200{,}000$ copies per time. The plotted interval is $0\leq Jt\leq15$; the simultaneous 99\% statement covers all 121 measured times through $Jt=30$. (b) Minimum over all edges of the exact witness value and the finite-record lower certificate; zero means inconclusive. (c) Exact first-return fidelity, the generator global-fidelity bound (clipped at zero), and the weakest-edge bound versus size. Return maxima are selected from the stated $3\leq Jt\leq7$ grid interval using noiseless dynamics, independently of readout records. (d) Empirical probability of certifying every edge at fixed $Jt=4.75$ in 400 independent readout experiments per cell. Each experiment has a 99\% simultaneous bound over its 20 sites, not over the entire displayed power scan. Binomial success counts and pointwise 95\% Wilson intervals for these power estimates are supplied in the data.}
\label{fig:certification}
\end{figure*}

We test the readout with $200{,}000$ fresh-copy weak outcomes at each of 121 fixed times $Jt=0,0.25,\ldots,30$, using $\mu=0.5$. A single $\alpha=0.01$ covers all 2,420 site--time intervals. At $Jt=4.75$, all 19 edges have positive lower certificates, with a minimum of $0.152389$ (Fig.~\ref{fig:certification}). Every interval in this realization contains its exact mean. The full record costs $24.2$ million independent preparations. This count is the sampling requirement in the simulated readout experiment; it refers neither to an experimentally collected dataset nor to a quantum state simulation with millions of qubits.

For the power calculations, we repeat the synthetic measurement experiment 400 times at each of 42 strength--budget combinations. As expected, a weaker detector needs more shots for certification. Five simultaneous interval failures occur in 16,800 independently generated experiments. These coverage events are consistent with the conservative $1\%$ family error allowance. The small observed failure count does not replace the concentration proof. The source includes data at every simulated time and the raw setting/outcome counts.

\subsection{Global-entanglement benchmarks with distinct measurement menus}
The generator sum is one of several global criteria. For the two-color partition $E\cup O$ of the path, the established two-setting witness uses $P_E=\prod_{i\in E}(\id+A_i)/2$ and $P_O=\prod_{i\in O}(\id+A_i)/2$~\cite{Toth2005,Zhou2019}. The target projector is $P_EP_O$, so $(\id-P_E)(\id-P_O)\geq0$ gives
\begin{equation}
 F_G\geq\langle P_E\rangle+\langle P_O\rangle-1
       \equiv L_2.
 \label{eq:twocolor}
\end{equation}
The exact first-return value is $0.508879$ at $N=18$ and $0.478850$ at $N=20$. It improves on $L_G$ and certifies the simulated $N=18$ state using exact expectations. We do not generate a finite-shot two-setting record. Each setting reads a full-register product basis and allows high-order correlations to be formed from resolved outcomes, unlike a single coarse local parity probe.

A complementary benchmark uses only bounded Pauli weight. For an open path, the restricted-measurement GME criterion of Li \emph{et al.}~\cite{Li2026} gives the sufficient condition
\begin{equation}
 \begin{split}
 \Delta_\gamma={}&\sum_i|\langle A_i\rangle|
  +\gamma\sum_{i=0}^{N-2}|\langle A_iA_{i+1}\rangle|\\
 &-\big[N-1+\gamma(N-2)\big]>0,\quad 0\leq\gamma\leq1.
 \end{split}
 \label{eq:lowweight}
\end{equation}
The absolute values make the criterion invariant under the known target signs. Adjacent products have weight four in the bulk and three at an end, so measuring both sums requires $2N-1$ bounded-weight observables. The margin is affine in $\gamma$; checking its two endpoints therefore gives the best margin in this family. Independent evaluations of the exact expectations at all seven first returns give $\Delta_0<0$ and $\Delta_1<0$. At $N=8$, the margins are $-0.006822$ and $-0.052918$; at $N=20$, they are $-1.160866$ and $-2.723305$. Table~S4 of the Supplemental Material gives the full size comparison~\cite{SupplementalMaterial}. On the $N=20$ time grid, $\Delta_1>0$ only at $Jt=0,0.05,0.10,0.15$, with no positive point at $Jt\geq3$.

These calculations use additional exact observables; the single-generator counts do not supply these certificates. Measuring the products in Eq.~\eqref{eq:lowweight} for a general state requires an enlarged menu. If exact dressed-blockade support is separately trusted, product means can be inferred from generator means. We evaluate them directly here, with the same reported negative margins. Failure at the returns does not show that GME is absent. Nor does the success of Eq.~\eqref{eq:edge} make the edge witness a stronger GME test, since the two tests address different entanglement properties.

\begin{figure*}[t]
\centering\includegraphics[width=\textwidth]{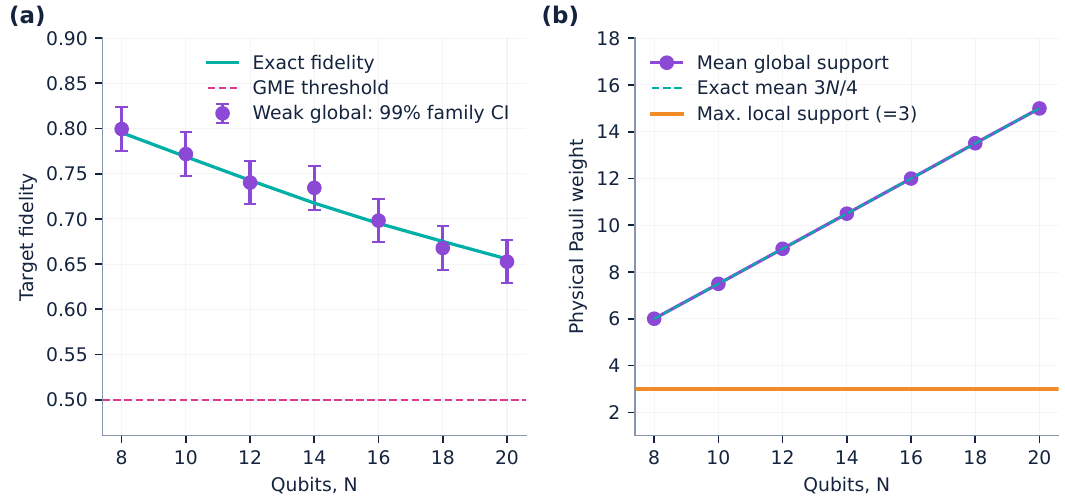}
\caption{A separate global certificate obtained by relaxing local support. (a) Ideal full-stabilizer-group weak readout at the first return, with $100{,}000$ fresh copies per size and contrast $\mu=0.5$. The displayed intervals have simultaneous 99\% coverage over the seven sizes. Every lower endpoint is above the connected-graph GME threshold $1/2$; at $N=20$ it is $0.6286064$. The cyan line is an exact simulation reference, while the purple intervals are computed from the sampled masks and outcomes. (b) Mean physical Pauli weight of those sampled settings, compared with its exact value $3N/4$ and the maximum weight three of the local menu. This comparator trades spatial support for global fidelity access. It assumes ideal global parity measurements; it is not included in the size-independent finite-pulse bound derived for a local generator. Different confidence families are used here and in the full time map of Fig.~\ref{fig:certification}, so their total-copy counts are not a matched-resource benchmark.}
\label{fig:sglobal}
\end{figure*}

Allowing broader measurement support permits global fidelity certification by uniform sampling of the full stabilizer group, as in direct fidelity estimation~\cite{Flammia2011,DaSilva2011}. One ideal weak group-parity check per fresh copy gives $\mathbb E[y/\mu]=F_G$. At the first return, $100{,}000$ copies per size and $\mu=0.5$ give simultaneous 99\% fidelity intervals across all seven sizes; every lower endpoint exceeds $1/2$ (Fig.~\ref{fig:sglobal}). At $N=20$ the interval is $[0.628606,0.676754]$, certifying GME independently of the local records. The sampled mean Pauli weight is $14.9957$, compared with at most three for the local checks. This straightforward global certificate uses additional spatial support. It gives neither a new direct-fidelity theorem nor a locality-preserving extension of Fig.~\ref{fig:certification}.

Table~S5 of the Supplemental Material compares the information, measurement support, and certified property for these different measurements~\cite{SupplementalMaterial}. Other bounded-weight or optimized global witnesses remain possible.

\subsection{Perturbations, alternative inputs and detector effects}
\begin{figure*}[t]
\centering\includegraphics[width=\textwidth]{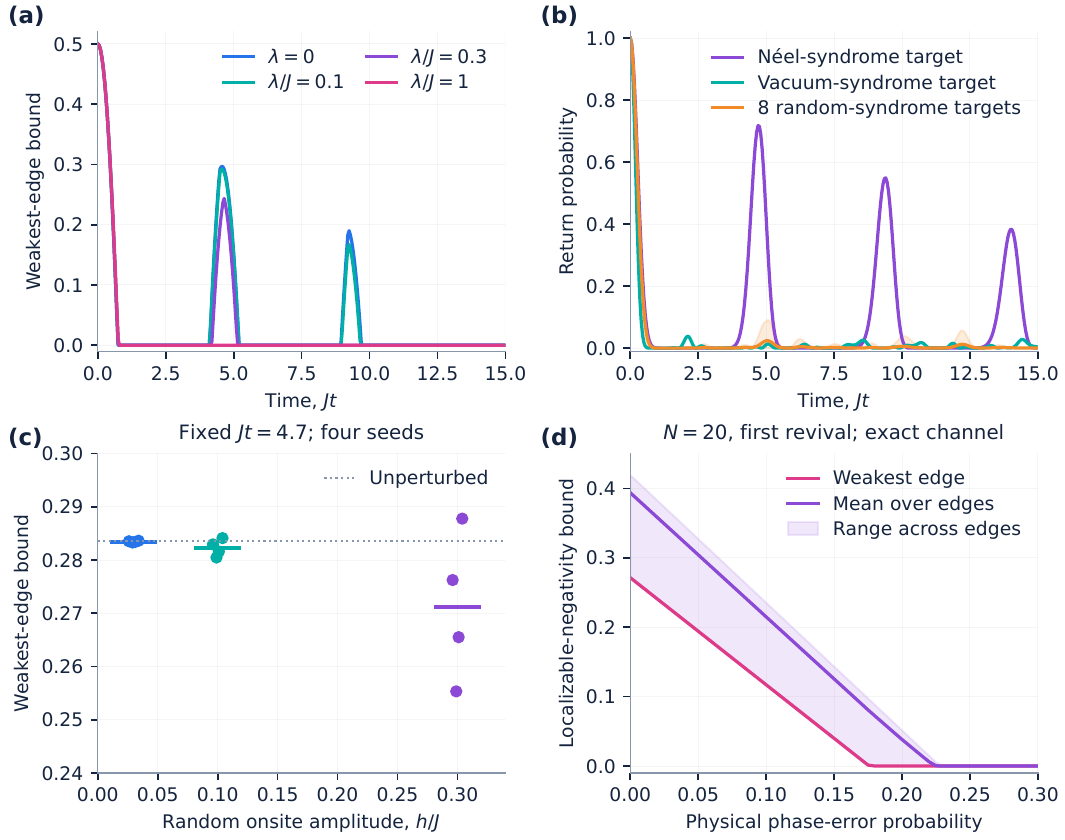}
\caption{Robustness and controls for the finite-time signal. (a) Exact weakest-edge witness for the $N=14$ N\'eel quench with a next-nearest-neighbor density interaction. (b) Return probabilities for the N\'eel, vacuum and eight random allowed reference inputs, all physically graph dressed; orange shading is their full realization range. (c) Weakest-edge bounds at the fixed time $Jt=4.7$ for four prescribed onsite-field realizations at each amplitude; short bars show the means and the dotted line shows the unperturbed value. These points are not confidence intervals. (d) Exact attenuation under independent physical $Z$ errors applied to the $N=20$ first-return state. The shaded band spans graph edges. The loss of a positive bound is a loss of this sufficient certificate, not proof of entanglement sudden death.}
\label{fig:robustness}
\end{figure*}

We test the reference-frame deformation
\begin{equation}
 H_\lambda=H_0+\lambda\sum_{i=0}^{N-3}n_in_{i+2}
                    +\sum_i h_i n_i .
 \label{eq:perturb}
\end{equation}
The physical version is $WH_\lambda W^\dagger$. These terms preserve blockade without requiring a target eigenstate; the initial variance stays at $NJ^2/2$. At $N=14$, $\lambda/J=0.1$ and $0.3$ retain a positive weakest-edge certificate at the first return, with exact bounds $0.277912$ and $0.232981$. At $\lambda/J=1$, the early return is substantially suppressed and the all-edge witness at its selected maximum is inconclusive (Fig.~\ref{fig:robustness}a). These few values cannot identify a critical coupling. The $\lambda/J=1$, $h_i=0$ point also needs care: its bulk coefficients satisfy the known hard-boson integrability condition $w^2=UV+V^2$ with $|w|=J$, $U=0$ and $V=\lambda$~\cite{Fendley2004,Turner2018b}. We use this point as a strong-deformation control, with no claim of robustness to generic nonintegrable perturbations; we have not established the boundary terms needed for an integrable finite open chain.

We dress control quenches from the vacuum and eight uniformly sampled allowed computational states with the same $W$, so every control also starts as a signed graph state. Their smaller early return probabilities show that initial graph entanglement alone does not ensure the N\'eel recurrence (Fig.~\ref{fig:robustness}b). Four fixed realizations at each of three onsite amplitudes give positive weakest-edge bounds at the fixed time $Jt=4.7$, including $h_i/J\in[-0.3,0.3]$ (Fig.~\ref{fig:robustness}c). We report the realization range; four samples do not constitute an asymptotic disorder ensemble.

Independent physical phase flips $Z_i$ with probability $p$ give $a_i\mapsto(1-2p)a_i$ exactly for any pre-noise state. Equation~\eqref{eq:edge} then gives the loss of localizable-entanglement certification at the first return (Fig.~\ref{fig:robustness}d). We evaluate these exact channel formulas on the simulated state, without a separate large-system density-matrix simulation. Symmetric classical outcome flips at known rate $e<1/2$ change the effective contrast to $\mu_{\mathrm{eff}}=(1-2e)\mu$, increasing the statistical cost after calibration. Unknown contrast or detector bias needs an independent bound before the certificate can be used.

\section{Finite-duration probes and energy-moment certification}\label{sec:finitepulse}
\subsection{A finite measurement pulse with the Hamiltonian active}
An ancilla initialized in $\ket0$, coupled through $A_i\otimes Y_a$ for duration $\tau$, and measured in $X_a$ realizes Eq.~\eqref{eq:instrument} in the impulsive limit. Write $\theta=\tfrac12\arcsin\mu$. For a finite pulse the exact unitary is
\begin{equation}
 U_\tau=\exp[-i(\tau H_G\otimes\id+\theta A_i\otimes Y_a)].
 \label{eq:gate}
\end{equation}
Let $C_i=\|[H_G,A_i]\|$. A unitary product-formula estimate gives
\begin{equation}
 \|U_\tau-e^{-i\tau H_G}\,e^{-i\theta A_i\otimes Y_a}\|
 \leq\frac{\tau\theta C_i}{2}.
 \label{eq:pulsebound}
\end{equation}
The displayed free unitary includes the ancilla identity and commutes with the ancilla readout. A sufficient systematic allowance in Eq.~\eqref{eq:interval} is therefore $b_i=\tau\theta C_i/\mu$. For the PXP model and the diagonal deformations in Eq.~\eqref{eq:perturb}, $C_i=2J$ on the blockade sector. The bound is independent of total system size and allows $H_G$ to remain active during measurement. Timing jitter and contrast uncertainty, if present, need separate additional allowances.

\begin{figure*}[t]
\centering\includegraphics[width=\textwidth]{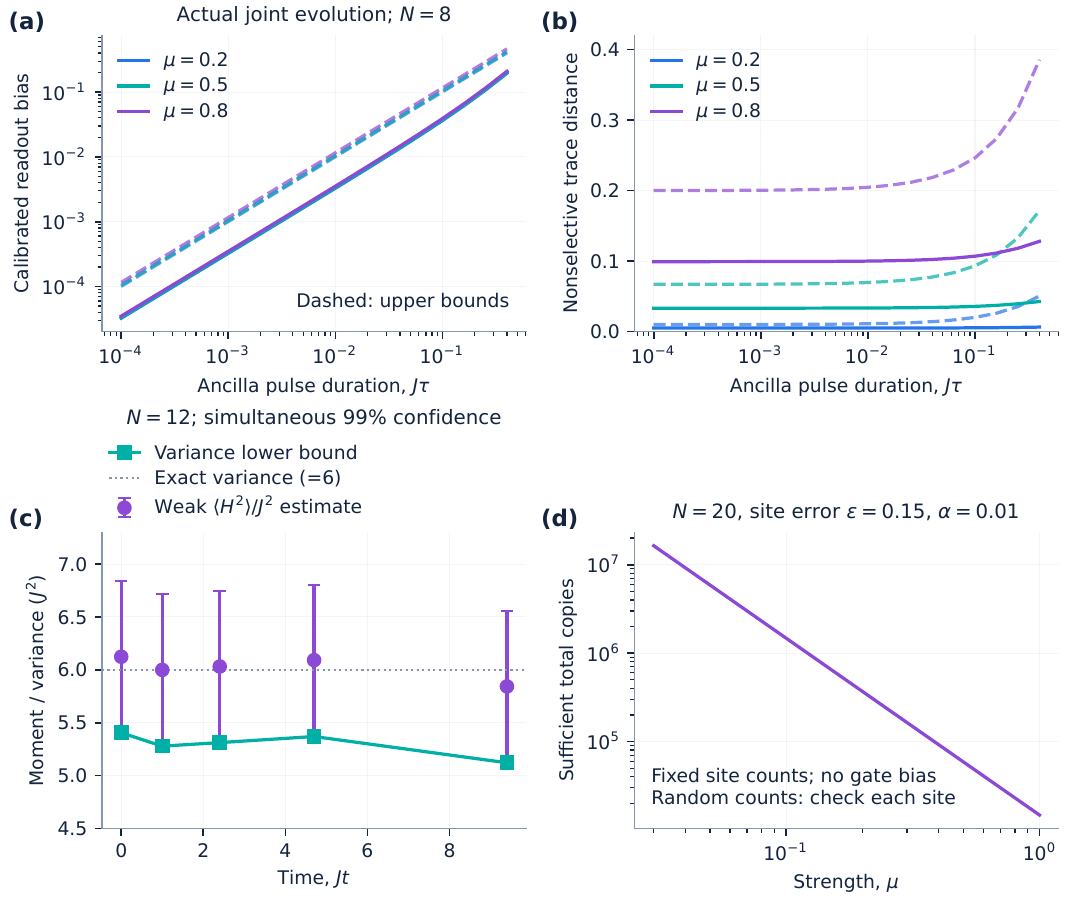}
\caption{Measurement dynamics and explicit resource costs. (a) Actual calibrated readout bias for a finite ancilla pulse on zero-based site 3 (the fourth qubit) of an $N=8$ snapshot at $Jt=1$, for three strengths; dashed curves show the calibrated bias bound $\tau\theta C_i/\mu$ implied by Eq.~\eqref{eq:pulsebound}. (b) Nonselective trace distance from free evolution and the corresponding bounds. Curves use joint matrix-exponential propagation, including $H_G$ throughout the pulse, represented in the equivalent reference frame. (c) Weak estimates of the energy second moment, their simultaneous confidence intervals, and propagated lower variance bounds for $N=12$. The exact mean energy is zero and exact variance is $6J^2$. (d) Analytic sufficient total copies for equal counts per site at one time with $\epsilon_{\mathrm{stat}}=0.15$, from Eq.~\eqref{eq:resources}; these are resource formulas, not additional quantum simulations.}
\label{fig:resources}
\end{figure*}

Combining the product-formula error with Eq.~\eqref{eq:channel} bounds the nonselective trace distance from the freely evolved input by $q+\tau\theta C_i/2$. Joint system--ancilla evolution at $N=8$ checks the calibrated mean-bias and disturbance inequalities for 57 duration--strength pairs (Fig.~\ref{fig:resources}). This finite-pulse validation uses eight sites; the commutator argument gives the analytic bound for larger chains, independently of their size. The local check is not quantum nondemolition with respect to $H_G$, since $[H_G,A_i]\ne0$ and the target wavepacket has positive energy variance.

The main $N=20$ time-map record uses the ideal instrument with $b_i=0$. For $\mu=0.5$ and $J\tau=0.01$, an implementation satisfying the model assumptions needs the extra allowance $b_i=0.010472$. Subtracting this allowance lowers an unclipped edge bound by at most that amount, less than the displayed ideal-record margin. This compares error budgets; it does not simulate a 20-spin finite-pulse readout record. A total disturbance cap must include $\tau\theta C_i/2$. Optimizing the statistical budget alone in Eq.~\eqref{eq:budgetcost} does not optimize the combined bias-and-variance error at finite duration. Sec.~S13 of the Supplemental Material gives the feasibility conditions for this combined budget~\cite{SupplementalMaterial}.

The fixed-allocation analysis can also explicitly account for total precision with a finite pulse. Let $a_\tau=\tau C_i/2=J\tau$ and let $\mu_{\max}$ be the largest strength satisfying the proven nonselective disturbance cap. Requiring the statistical radius plus worst-case bias to be at most $\epsilon_{\mathrm{tot}}$ replaces $\epsilon_{\mathrm{stat}}\mu$ in the cost by
$g_\tau(\mu)=\epsilon_{\mathrm{tot}}\mu-a_\tau\arcsin\mu$.
A finite sufficient budget needs $\epsilon_{\mathrm{tot}}>a_\tau$. When this holds, maximizing $g_\tau$ over $0<\mu\leq\mu_{\max}$ gives
\begin{equation}
 \mu_{\mathrm{tot}}=\min\!\left\{\mu_{\max},
 \sqrt{1-\left(\frac{a_\tau}{\epsilon_{\mathrm{tot}}}\right)^2}\right\}.
 \label{eq:totalprecisionstrength}
\end{equation}
Here, $g_\tau'(\mu)=\epsilon_{\mathrm{tot}}-a_\tau/\sqrt{1-\mu^2}$, and the sufficient cost is $2N\log(2N/\alpha)/g_\tau(\mu)^2$ whenever $g_\tau>0$. The upper contrast bound follows from $q+(a_\tau/2)\arcsin\mu\leq\delta$. With finite-pulse bias included, the largest contrast allowed by disturbance need not minimize the total-precision cost. This result minimizes the displayed conservative budget for the fixed instrument and allocation. It does not optimize actual state-dependent measurement error or all possible protocols.

\begin{figure*}[t]
\centering\includegraphics[width=0.94\textwidth]{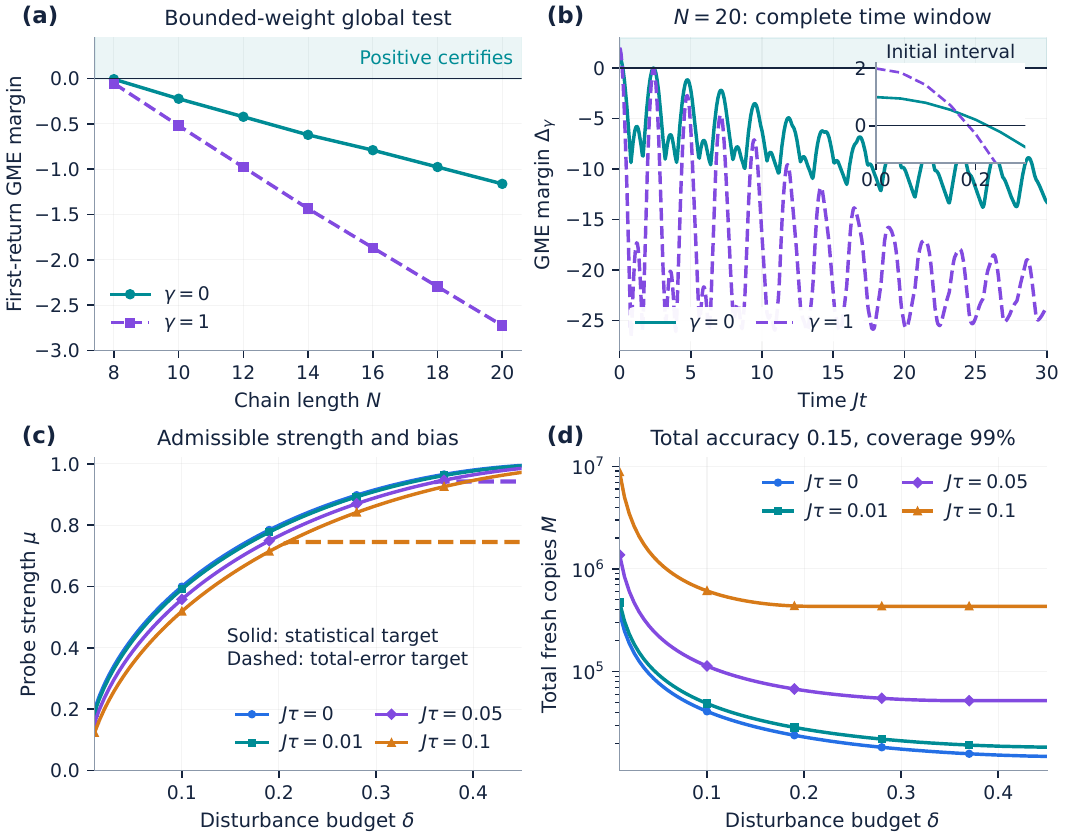}
\caption{Additional global tests and a disturbance-constrained probe design. (a) Exact-expectation GME margins of the bounded-weight path criterion at the first return for seven sizes. A positive margin is sufficient; neither endpoint of the one-parameter family certifies these returns. (b) The complete $N=20$ time trace for the same two margins. The shaded positive half-plane is the certification region; the inset resolves the initial interval. (c) Largest strength allowed by the sufficient finite-pulse disturbance bound for four pulse durations. Solid curves optimize a fixed statistical radius. Dashed curves also account for bias at fixed total accuracy, and can therefore saturate earlier. (d) Integer total preparation budgets for the total-accuracy design. Panels (c,d) use $N=20$, one predeclared time, $\alpha=0.01$ and $\epsilon_{\rm tot}=0.15$. Each point assigns an equal integer number of independent preparations to every generator. The curves are evaluated bounds, not empirical performance or hardware data, and do not imply optimality over other measurement protocols.}
\label{fig:srevision}
\end{figure*}

Figure~\ref{fig:srevision}(c,d) gives the resulting strength and preparation budgets; panels (a,b) show the bounded-weight global margins of Sec.~\ref{sec:benchmarks}. These curves evaluate the resource formulas, separately from the finite-gate simulations in Fig.~\ref{fig:resources}.

\subsection{An independent certificate of nonzero energy variance}
Sparse Pauli sampling of $H_G$ and $H_G^2$ also certifies nonzero energy variance without reconstructing the state. At $N=12$, two million shots per observable at each of five fixed times, with $\mu=0.5$, give a simultaneous 99\% lower variance bound above $5.1211J^2$, compared with the exact $6J^2$. The energy-term means remain exactly blind, while the second moment establishes that the prepared cluster trajectory is not an eigenstate. We account separately for the second-moment menu and its larger measurement support below.

For energy measurements, write $H_G=c_0\id+\sum_\ell c_\ell B_\ell$ in distinct Hermitian Pauli strings. Draw a nonidentity term with probability $|c_\ell|/B$, $B=\sum_{\ell\neq0}|c_\ell|$, and report $B\operatorname{sgn}(c_\ell)y/\mu+c_0$. Its expectation is the energy, with Hoeffding radius $B\sqrt{2\log(2/\alpha_H)/(M\mu^2)}$. To expand $H_G^2$, we multiply and collect Pauli strings algebraically; anticommuting cross terms cancel, and we add the known identity coefficient without sampling. Products can span separated regions and have larger support than a local three-site stabilizer, so the second-moment branch is not assigned the same local gate cost. Ten independent expectation estimates (two observables at five times) each use $\alpha_H=0.001$, giving a total error probability of at most 0.01. The variance lower bound is
\begin{equation}
 \operatorname{Var}_\rho(H_G)\geq\big[\widehat{H^2}-r_{H^2}-(|\widehat H|+r_H)^2\big]_+ .
 \label{eq:variancecertificate}
\end{equation}
This inequality holds on the simultaneous moment-confidence event. It follows from $|\langle H_G\rangle|\leq|\widehat H|+r_H$; squaring an unbiased energy estimate does not give an unbiased estimate of the squared mean energy. At $N=12$, the Pauli expansions contain 44 nonidentity terms for $H_G$ and 754 for $H_G^2$; the known identity coefficient of $H_G^2$ is $3.5J^2$. The five-time energy branch needs 20 million additional preparations. A positive variance rules out support entirely in a single energy eigenspace, but alone certifies neither scarring nor entanglement.

\section{Numerical implementation and independent validation}\label{sec:numerics}
\subsection{State representation, dynamics and spectra}
We set $\hbar=J=1$ in all simulations, with unit coefficient for $PXP$. Computational bit $i$ is the $i$th least significant bit. Numbered data plots label vertices $1,\ldots,N$; equations and code use $0,\ldots,N-1$. The allowed basis states are sorted integers $b$ satisfying $b\mathbin{\&}(b\!\gg\!1)=0$. The sparse Hamiltonian connects $b$ to $b\oplus2^i$ only when both existing neighbors of $i$ are empty. We evaluate diagonal terms directly on these bitstrings. The largest dynamical calculation has $N=20$ and dimension 17,711. SciPy's matrix-exponential action propagates the state at 601 uniform times from zero to 30, without Trotter steps or tensor-network truncation~\cite{AlMohy2011,Virtanen2020}. Across the main matrix-exponential dynamical suite, the maximum recorded norm error is below $5.1\times10^{-14}$ and the maximum energy drift below $3.9\times10^{-13}J$.

We define the first return as the largest local return-probability maximum on the grid within $3\leq Jt\leq7$. This rule uses noiseless evolution, independently of the noisy certification record, and identifies a finite-time peak rather than an asymptotic revival period. For spectra, normalized symmetric and antisymmetric reflection-orbit vectors form separate isometries. We use dense Hermitian diagonalization within each sector for $N=10,12,14,16$, with additional $\lambda/J=0.3$ samples at $N=12,14$. Eigenpair residuals stay below $5.2\times10^{-14}J$, and the largest orthogonality error is $2.3\times10^{-12}$.

We compute gap ratios after removing $|E|\leq10^{-9}J$, using the central half by index of the remaining spectrum in each reflection sector. Triplets crossing zero are excluded only when zero modes have been removed; spacings below $10^{-10}J$ would also be excluded and counted. We report this zero-specific convention because protected degeneracies would otherwise dominate the statistic; the threshold is not fitted to obtain a target ratio. In the undeformed model, chiral symmetry also correlates the positive- and negative-energy ratios. The deterministic size series gives no independent disorder samples, so we assign no bootstrap confidence interval to correlated neighboring ratios.

To calculate physical entropies, we embed the constrained vector in the full $2^N$ computational basis, apply a normalized Walsh--Hadamard transform and open-chain CZ phases, and find singular values across the $N/2$ cut. We express entropies in bits and evaluate every eigenvector for $N\leq14$. At $N=16$, the entropy sample contains the $N+1$ largest nonzero N\'eel-overlap eigenvectors and 160 uniformly chosen remaining nonzero-energy vectors (seed 72391). The plots and captions specify this selection. We sum zero-mode overlaps over the entire degenerate subspace. The random-state Page value is not an assumed exact entropy reference for a constrained system at finite energy~\cite{Page1993}.

\subsection{Synthetic records and comparison observables}
We generate synthetic weak records directly from the Born probabilities of Eq.~\eqref{eq:instrument}. A multinomial draw determines the site counts, and independent conditional binomial draws determine the positive outcomes. These draws retain the sufficient statistics of independent shots and introduce no approximation to the measurement distribution. Seed 870031 sets the $N=20$ record, 870032 the 400-repetition power experiments, and 870033 the energy records. For the time map, the simultaneous family contains all $20\times121$ intervals at $\alpha=0.01$. Each power experiment uses its own 20-site family at the same $\alpha$. Pointwise 95\% Wilson intervals describe uncertainty in the empirical success proportions, rather than in the quantum means.

For a uniform binary mask $r$, the signed group element is $S_r=\prod_iA_i^{r_i}$. The masks include the identity, and each setting uses the ideal weak effect $(\id+y\mu S_r)/2$. We obtain its exact mean from the Walsh transform of the reference-frame probability distribution, then sample the binary outcome. The identity $2^{-N}\sum_r S_r=\ket{G_s}\bra{G_s}$ gives the unbiased estimator $\widehat F=M^{-1}\sum y/\mu$. For $K=7$ sizes, the simultaneous radius is $\sqrt{2\log(2K/\alpha)/(M\mu^2)}=0.0240736$ with $M=10^5$, $\mu=0.5$, $\alpha=0.01$. The seeds are $870034+N$. We retain each group element's mask, outcome and physical Pauli weight; the latter has exact mean $3N/4$ on this path graph. Direct physical stabilizer actions independently check all 256 group means at $N=8$. This comparison uses ideal global parities and falls outside the local finite-pulse resource guarantee.

In the bounded-weight comparison, independently integrated reference-frame states give $\langle A_iA_{i+1}\rangle=s_is_{i+1}\langle Z_iZ_{i+1}\rangle$. We select first-return points by the same noiseless rule as in Fig.~\ref{fig:certification}c. A separate $N=20$ integration covers all 601 times; its maximum stabilizer and fidelity discrepancies from the main exponential propagation are $7.55\times10^{-12}$ and $1.71\times10^{-12}$. This calculation checks Eq.~\eqref{eq:lowweight} without generating further weak outcomes. The source package includes the parameters, all seven size records, and the full time trace.

\subsection{Perturbation ensembles and independent verification}
We scan interactions at $\lambda/J=0,0.1,0.3,1$. Random initial allowed bitstrings use seeds 117, 229, 331, 443, 557, 661, 773 and 887. The onsite coefficients are drawn independently and uniformly from $[-h,h]$ at $h/J=0.03,0.1,0.3$, with seeds 31, 73, 137 and 211. We retain every realization regardless of its observed spectrum or return. All random draws use NumPy's documented default generator~\cite{Harris2020}.

The supplied verification script builds full Kronecker-product Hamiltonians independently of the production transition builder for $N=4,5,6,7$. It checks dense exponential evolution, the explicit Clifford circuit, Pauli expansions of $H$ and $H^2$, target variance, Hamiltonian-term blindness and the global projector inequality. Direct $Z$ localization tests cover 900 mixed-state/edge combinations drawn from noisy graph, random mixed and fully separable inputs. Explicit Kraus matrices independently check the weak effects, averaged channel and tight disturbance bound. All 86 named checks meet their recorded tolerances. These checks test implementations of mathematical statements and finite-size calculations; they do not provide formal proofs or establish thermodynamic thermalization. A further audit constructs the Hamiltonian from tensor products and integrates eight trajectories through $N=14$ with DOP853 at two tolerances. At relative tolerance $10^{-12}$, the largest stabilizer discrepancy from exponential propagation is $2.4\times10^{-12}$. The source package also records state-independent finite-pulse operator checks and exact binomial-tail checks.

The finite-gate calculation propagates the joint ancilla and an $N=8$ PXP snapshot at $Jt=1$, on zero-based site 3, with strengths $0.2,0.5,0.8$ and 19 logarithmically spaced durations from $10^{-4}/J$ to $0.4/J$. We compare the reduced output density matrix with free evolution by diagonalizing their Hermitian difference. Unitary covariance gives the graph-frame result. A local Qiskit implementation independently checks full-space Pauli assembly, physical Clifford circuits and 81 weak-ancilla gate cases for $N=4,6,8$. The maximum statevector discrepancy is $9.5\times10^{-15}$ and the maximum weak-readout mean error is $6.7\times10^{-16}$. Qiskit exact synthesis also uses SciPy internally: the construction and circuit representation are independent, but the underlying linear-algebra library is shared. The reproducibility records preserve its environment and the audit of an initially failed script. We generate plots from saved arrays using Matplotlib and supply the vector PDF/SVG originals~\cite{Hunter2007}.

\section{Discussion and outlook}\label{sec:discussion}
We give a measurement protocol for entanglement in approximate scarred dynamics. It combines a measurement-menu blindness result, established graph-edge negativity witnesses, simultaneous finite-record confidence bounds, and finite-pulse error and disturbance estimates under the stated measurement assumptions. In the simulations, a known PXP quench in a physically graph-entangled frame retains certifiable localizable entanglement at imperfect revivals across the tested sizes and perturbations; this does not establish a new scar phase.

The graph dressing creates the initial cluster structure; scarring explains its recovery. For comparison, the integrable Hamiltonian $W(J\sum_iX_i)W^\dagger=J\sum_iZ_i$ also returns $W\ket{\Ztwo}$ exactly at multiples of $\pi/J$. Recurrence and graph-edge positivity alone establish neither nonintegrability nor scarring. We also use a specified nonintegrable model, symmetry-resolved spectra, exceptional overlaps and alternative-input controls. A local witness cannot reconstruct an unknown arbitrary Hamiltonian's spectrum. A recent proof excludes nontrivial finite-range local conserved charges for periodic PXP under stated locality and blockade-sector conditions~\cite{Park2025}. This supports the nonintegrable setting, but proves neither thermodynamic ETH nor nonintegrability of every finite open-chain deformation studied here.

A positive local certificate remains useful when a chosen global witness fails, without making it a better GME detector. The generator sum, two-color projectors, bounded-weight adjacent products, and full-group fidelity record access different information. Their first-return outcomes set the scope of this benchmark; none represents all economical global tests~\cite{Toth2005,Toth2005b,Zhou2019,Li2026}. We certify a graph-edge resource in the recurrent wavepacket, without claiming that an individual PXP eigenvector is a cluster state.

The protocol estimates the pre-check state from independent, identically prepared snapshots at each designated time. It neither prepares a cluster state autonomously nor treats consecutive readouts of one disturbed trajectory as independent samples. Continuous stochastic evolution needs a specified instrument or unraveling, even with an unconditional Lindblad generator~\cite{Lindblad1976,Gorini1976,Dalibard1992,Brun2002,Jacobs2006}. The confidence guarantee excludes adversarially correlated preparations, adaptive stopping and temporal drift. Martingale methods address other sequential sampling questions, but do not by themselves recover an unmonitored state from a monitored trajectory~\cite{Azuma1967,Freedman1975}. Sampling without replacement also needs a different sampling theorem~\cite{Serfling1974}.

A per-copy disturbance budget gives a reason to use a weak probe. For the specified ideal square-root instrument and equal fixed allocation, Eq.~\eqref{eq:budgetcost} gives the strength minimizing the Hoeffding sufficient cost at fixed statistical accuracy. Other instruments, state-specific error criteria, or total finite-pulse error optimization may favor different strengths. Any practical comparison must also count the costs of implementing $W$, routing the ancilla, and repeatedly preparing the state on the same device.

This finite-size benchmark identifies uninformative Hamiltonian readouts, local observables that certify recoverable edge entanglement, and measurements that support stronger global conclusions. Unknown Hamiltonians or certification of entangled scar eigenstates would require further identification or spectral-resolution assumptions. The present formulas give confidence guarantees for a specified model and instrument, without solving that broader reconstruction problem.

\begin{acknowledgments}
\noindent\textit{Acknowledgments and funding:}\par
\noindent\rule{0.95\columnwidth}{0.35pt}
\end{acknowledgments}

\section*{Author contributions}
\noindent\rule{0.95\columnwidth}{0.35pt}

\section*{Competing interests}
\noindent\rule{0.95\columnwidth}{0.35pt}

\section*{Data availability}
The data underlying the figures are supplied with the accompanying source package in NPZ and CSV form, including weak-measurement setting/outcome counts, prescribed seeds, parameters, and validation records. The records are synthetic; no experimental observations are reported. The power scan includes aggregate success and coverage-failure counts and scripts for regenerating individual trials.\par
\noindent\textit{Public repository identifier:}\ \rule{0.49\columnwidth}{0.35pt}

\section*{Code availability}
The accompanying package contains the article and Supplemental Material sources, bibliography, simulation and validation code, figure scripts, environment requirements, and rebuild instructions.\par
\noindent\textit{Public code repository identifier:}\ \rule{0.39\columnwidth}{0.35pt}

\bibliographystyle{apsrev4-2}
\makeatletter\immediate\write\@auxout{\string\citation{apsrev42Control}}\makeatother
\bibliography{aps_control,references}
\end{document}